\documentclass{osa-article}

\journal{oe}

\usepackage{lineno}
\usepackage{booktabs}
\usepackage{bm}
\usepackage{subfig}

\begin{document}

\title{Robust photon-efficient imaging using a pixel-wise residual shrinkage network}

\author{Gongxin Yao\authormark{1,$\dagger$}, Yiwei Chen\authormark{1,$\dagger$}, Yong Liu \authormark{1}, Xiaomin Hu \authormark{2,3}, and Yu Pan\authormark{1,*}}

\address{\authormark{1} State Key Laboratory of Industrial Control Technology, Institute of Cyber-Systems and Control, \\ \,\,\,\,College of Control Science and Engineering, Zhejiang University, Hangzhou, 310027,\\ \,\,\,\,People’s Republic of China\\
}

\address{\authormark{2} CAS Key Laboratory of Quantum Information, University of Science and Technology of China, Hefei,\\ \,\,\,\,230026, People’s Republic of China\\
}

\address{\authormark{3} CAS Center For Excellence in Quantum Information and Quantum Physics, University of Science and\\ \,\,\, Technology of China, Hefei, 230026, People’s Republic of China
}

\address{\authormark{$\dagger$} These authors contributed equally to this work\\
}

\email{\authormark{*}ypan@zju.edu.cn} 



\begin{abstract}
Single-photon light detection and ranging (LiDAR) has been widely applied to 3D imaging in challenging scenarios. However, limited signal photon counts and high noises in the collected data have posed great challenges for predicting the depth image precisely. In this paper, we propose a pixel-wise residual shrinkage network for photon-efficient imaging from high-noise data, which adaptively generates the optimal thresholds for each pixel and denoises the intermediate features by soft thresholding. Besides, redefining the optimization target as pixel-wise classification provides a sharp advantage in producing confident and accurate depth estimation when compared with existing research. Comprehensive experiments conducted on both simulated and real-world datasets demonstrate that the proposed model outperforms the state-of-the-arts and maintains robust imaging performance under different signal-to-noise ratios including the extreme case of 1:100.
\end{abstract}


\section{Introduction}

Active optical imaging technology has greatly expanded the boundaries of human vision and achieved rapid progress in many emerging applications, such as long-range imaging \cite{pawlikowska2017single, chan2019long}, underwater imaging \cite{halimi2017object, maccarone2015underwater}, non-line-of-sight imaging \cite{o2018confocal, liu2019non, saunders2019computational} and microscope imaging \cite{bruschini2019single, casacio2021quantum}. In recent years, light detection and ranging (LiDAR) \cite{schwarz2010mapping} has secured a dominant position in the field of active optical imaging. Conventional LiDAR system requires more than $10^{3}$ signal photons per pixel for accurately imaging a 3D scene \cite{shin2015photon}. However, such large number of signal photons can not be collected in photon-starved scenarios where optical flux and integration time are limited. Since single-photon avalanche diodes (SPADs) \cite{rochas2003single} provide extraordinary optical sensitivity and picosecond ($ps$) time resolution, SPAD-based LiDAR, also named single-photon LiDAR, has been designed to fully exploit few or even one signal photon per pixel to precisely capture the 3D structure information in photon-starved scenarios, such as extremely distant and poorly reflective surfaces \cite{pellegrini2000laser, mccarthy2013kilometer, pawlikowska2017single}.

\begin{figure}[t]
	\begin{center}
		\includegraphics[width=0.85\linewidth]{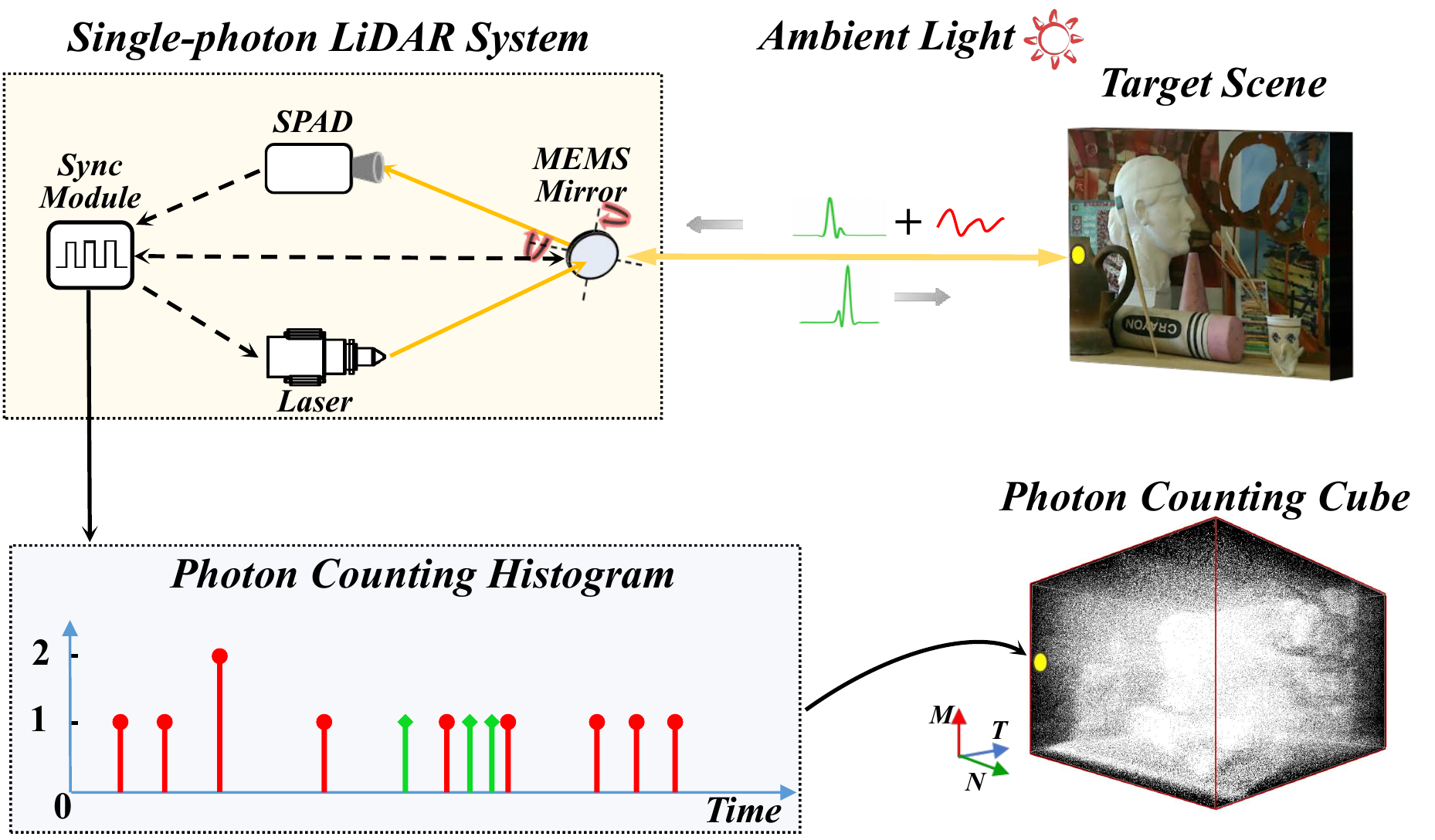}
	\end{center}
	\vspace{-6mm}
	\caption{A typical coaxial single-photon LiDAR system. The laser emits periodic pulsed lights towards an MEMS mirror which redirects the pulses to the target. The synchronization module acts as a central controller, which receives the signals from the SPAD and records the time of flight (ToF) between the target and detector. Based on repeated measurements, a photon counting histogram can be built. Signal and noise photons are marked in green and red, respectively.}
	\label{fig:Lidar}
\end{figure}

A typical coaxial single-photon LiDAR system as shown in Fig.~\ref{fig:Lidar} consists of a pulsed laser, a Micro-Electro-Mechanical system (MEMS) mirror, a SPAD sensor and the synchronization module. In practice, the recorded data may contain a large number of false counts due to the strong ambient lights and dark counts of SPAD. Thus, a key challenge is to robustly recover the precise depth images from such high-noise counting data. Many efforts have been devoted to developing statistical-learning-based denoising algorithms. For example, Shin \textit{et al.} \cite{shin2015photon, shin2016photon} proposed a constrained maximum likelihood (ML) estimation algorithm with a total variance regularization. Following this line, Rapp \textit{et al.} \cite{rapp2017few} designed a cluster searching strategy and exploited the estimated reflectivity to improve the performance of the constrained ML estimation. However, these statistical approaches often suffer from performance degradation as the background noise increases. 

In recent years, deep learning approaches \cite{lindell2018single, peng2020photon, zang2021non, usmani2021three} have demonstrated superior performance for photon-efficient imaging in high-noise scenarios. Lindell \textit{et al.} \cite{lindell2018single} proposed a network based on U-Net \cite{ronneberger2015u} for imaging at low signal-to-background ratios (SBRs) with sensor fusion strategy. A similar deep sensor fusion strategy that combines RGB images and corrupted SPAD data to estimate the depth of a scene is also proposed in \cite{sun2020spadnet}. However, these approaches require additional intensity images which are difficult to obtain in realistic scenarios. Peng \textit{et al.} \cite{peng2020photon} proposed a non-local neural network to extract the long-range correlations among the photon counting data across the whole temporal dimension. However, the large computational and memory consumption of the non-local block may reduce the efficiency of this model. Based on U-Net++ \cite{zhou2019unet++}, Zang \textit{et al.} \cite{zang2021non} built a network with rich skip connections between multi-scale features to capture the spatial patterns, while such complex connections are more compatible with clean data. Most recently, Zhao \textit{et al.} \cite{zhao2022photon} also adopted the U-Net++ architecture and employed an Attention-Directed Attention Gate module to enhance the edges' reconstruction. Generally speaking, existing deep learning models focus more on capturing spatial or temporal correlations from the data while lack a careful consideration for denoising. As a result, the robustness against noise cannot be guaranteed, i.e., these algorithms may fail to behave stably at different SBRs. For this reason, the goal of this paper is to design a denoising network to achieve robust imaging performance at different SBRs.

Soft thresholding has been widely used as an efficient denoising approach in signal processing \cite{daubechies2009wavelet, donoho1994ideal} and recently has drawn attention from deep learning communities \cite{isogawa2017deep, zhao2019deep}. The core idea behind soft thresholding is to use thresholds to distinguish real signals from noises, while the input signals below the thresholds are zeroed and the others are shrinked. Since determining the task-specific thresholds requires significant expertise, Isogawa \textit{et al.} \cite{isogawa2017deep} proposed a neural network to learn the global thresholds for all inputs, and the learned values are fixed after training. Zhao \textit{et al.} \cite{zhao2019deep} designed a residual shrinkage network to produce dynamic thresholds for different inputs. In particular, the thresholds are designed channel-wise and shared among all pixels, and then soft thresholding shrinkage is leveraged to denoise the data channel by channel. However, since the data collection with respect to each pixel can be treated as an independent process, the optimal thresholds for each pixel must be different from each other. Therefore, the global or channel-wise threshold may not be the best solution for denoising the photon counting data.

In this paper, we propose a pixel-wise residual shrinkage block (PRS block) to improve the feature denoising ability from high-noise SPAD data, with the ultimate goal of generating precise depth images at different SBRs. PRS block calculates the dynamic threshold for each pixel of the input and implements soft thresholding as a nonlinear transformation layer to increase the SBR. The PRS-Net is built by stacking multiple PRS blocks. In addition, an improved temporal window strategy based on a specific 3D convolution is proposed to cluster the signal photons before the PRS blocks. Another advantage brought by the pixel-wise approach is that we can redefine the loss function with cross entropy to improve its optimization performance. More specifically, the previous methods tried to minimize the reconstruction error between the distribution of predictions and a predetermined waveform based on Kullback-Leibler $(\mathrm{KL})$ divergence, where the predetermined waveform is generated from the real depth of each pixel. However, the real depth must be located within one time bin, and thus using a specific pulse waveform that spans a number of time bins to replace the real label is inherently inaccurate. Moreover, since the cross entropy loss can produce a more centralized prediction with the accurate label, the approximation to total variation loss also becomes more accurate than the previous methods. Experiments on simulated dataset demonstrate that PRS-Net achieves better performance than the state-of-the-art approaches, especially in extremely low SBRs. Moreover, on the real-world dataset captured in \cite{lindell2018single}, PRS-Net outperforms the state-of-the-arts by a large margin.


\section{Preliminary}
\subsection{Problem definition}
We assume that the pulse repetition period is long enough to avoid the distance aliasing problem \cite{shin2015photon}. Denote the total number of time bins as $T$, and the photon counting histogram $h_{ij} \in \mathbb{R}^{T}$ is recorded for each pixel. The photon counting cube $H \in \mathbb{R}^{T \times M \times N}$ is obtained by combining all the counting histograms, where $M \times N$ is the resolution of the LiDAR system. Under the single-surface assumption \cite{shin2016photon}, $H$ is used to recover a depth image $\hat{Z} \in \mathbb{R}^{M \times N}$ by
\begin{equation}
    \hat{Z} = f(H),
\end{equation}
where $f(\cdot)$ represents the reconstruction algorithm. The key challenge of this task is to recover a faithful depth image from the counting cube with limited signal photon counts. Ideally (with no background count), it may be possible to search for the time bin with most photon counts for the estimation of the depth. However in practice, the recorded histogram may contain a significant number of background counts, which would make real signals indistinguishable from noise. In this case, the aim is to denoise these fake counts and generate $\hat{Z}$ as close as possible to the ground-truth depth image $Z$.

\subsection{Observation model}
We briefly review the observation model in \cite{shin2015photon} for generating the simulated dataset for training. Denote the impulse response function of the pulsed light as $s(t)$, and $n_{b}$ denotes the average noise intensity including the background light and dark count of SPAD. For the ($i,j$)-th pixel, the photon detections produced by SPAD can be modelled by an inhomogeneous Poisson process \cite{snyder2012random} with the time-varying rate function as
\begin{align} 
    \lambda_{ij}(t) = \eta \cdot \alpha_{ij} \cdot s(t - \frac{2Z_{ij}}{\mathrm{c}}) + n_{b},
\end{align}
where $\mathrm{c}$ is the speed of light and $\alpha_{ij}$ is the light energy attenuation ratio due to atmospheric attenuation and diffuse reflection. $\eta$ is the quantum efficiency of the detector. Based on the rate function, the expectation value of the detected photon counts at the ($i,j$)-th pixel during an integration period $\Delta$ is calculated by
\begin{align} 
    \mathrm{E}_{ij} (T_{b}, T_{b} + \Delta) = \int_{T_{b}}^{T_{b} + \Delta} \lambda_{ij}(t)dt,
\end{align}
where $T_{b}$ denotes the beginning time of the integration period. Since the detection of each photon is independent of each other, the photon counting histogram of the ($i,j$)-th pixel based on $N$ illuminations can be sampled by following Poisson process as
\begin{equation}
    \label{Eq:poisson}
    h_{ijk} \sim \mathrm{Poisson}(N \cdot \mathrm{E}_{ij}(k\Delta,(k+1)\Delta)), \quad k = 1,2 \cdots,T.
\end{equation}
Based on the observation model, simulated photon counting cubes can be sampled using the depth images and reflectivity maps.

\begin{figure}[t]
	\centering
	\includegraphics[width=\linewidth]{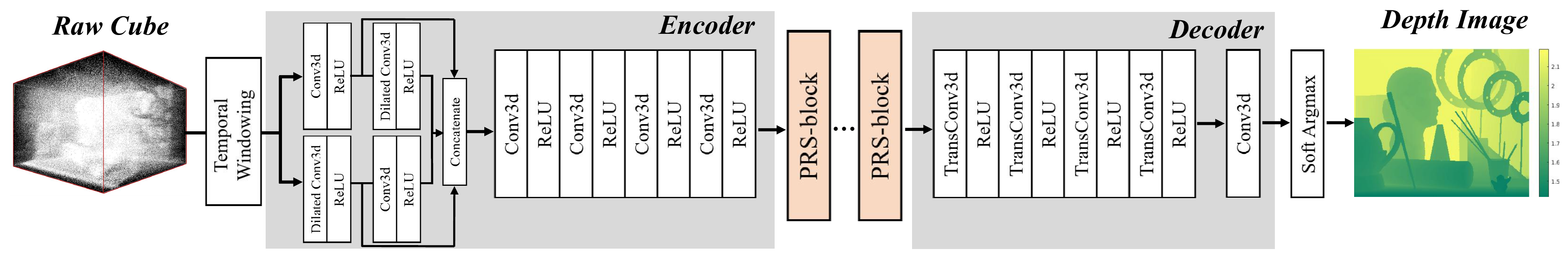}
	\caption{The overall structure of the PRS-Net.}
	\label{fig:overview}
\end{figure} 


\section{Approach}
In this section, we describe the structure of the PRS-Network and its optimization target which redefines the imaging task as a pixel-wise classification problem.

\subsection{Network structure}
Figure \ref{fig:overview} shows the overall structure of the PRS-Net, which consists of a temporal windowing module, an encoding module, a series of PRS blocks and a decoding module.

\subsubsection{Temporal windowing module}
\begin{figure}[t]
	\centering
	\subfloat[]{\includegraphics[width=.45\linewidth]{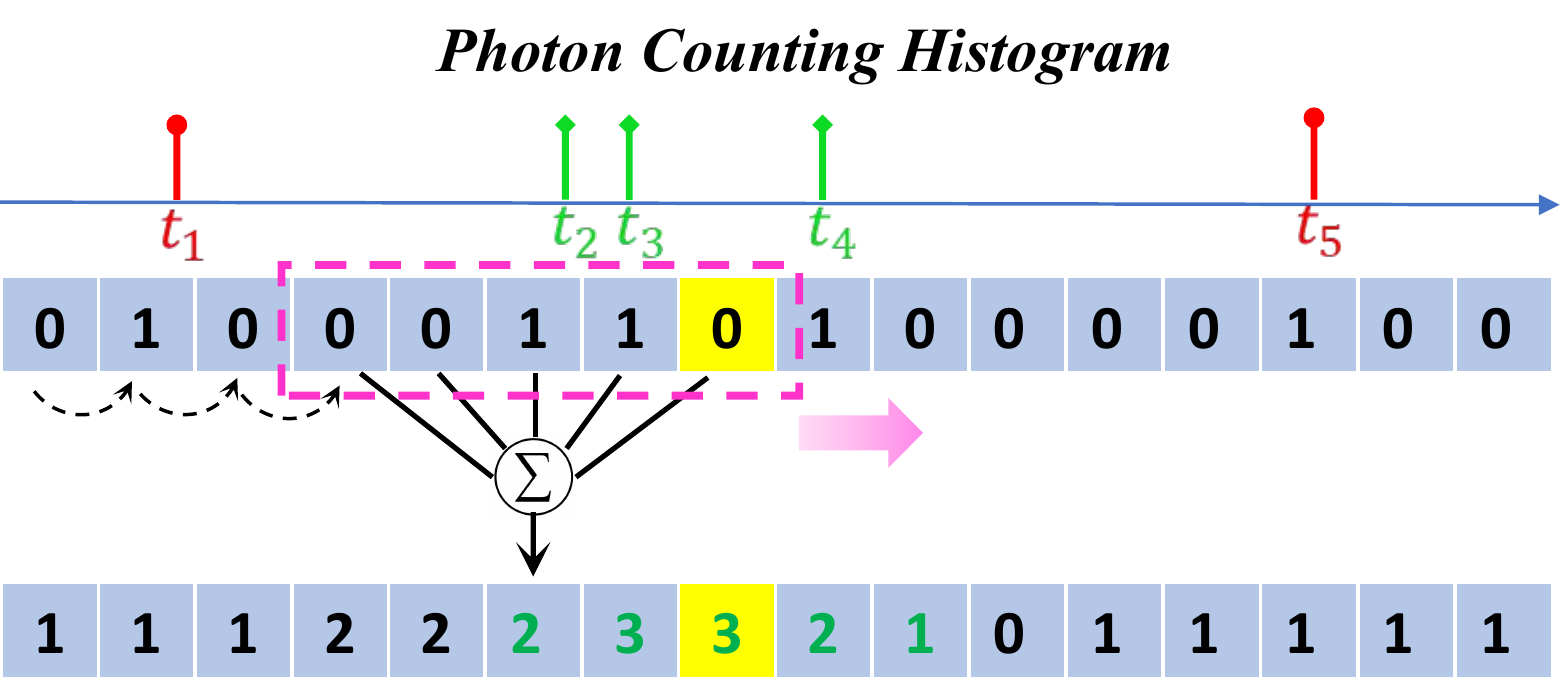}}
	\quad
	\subfloat[]{\includegraphics[width=.45\linewidth]{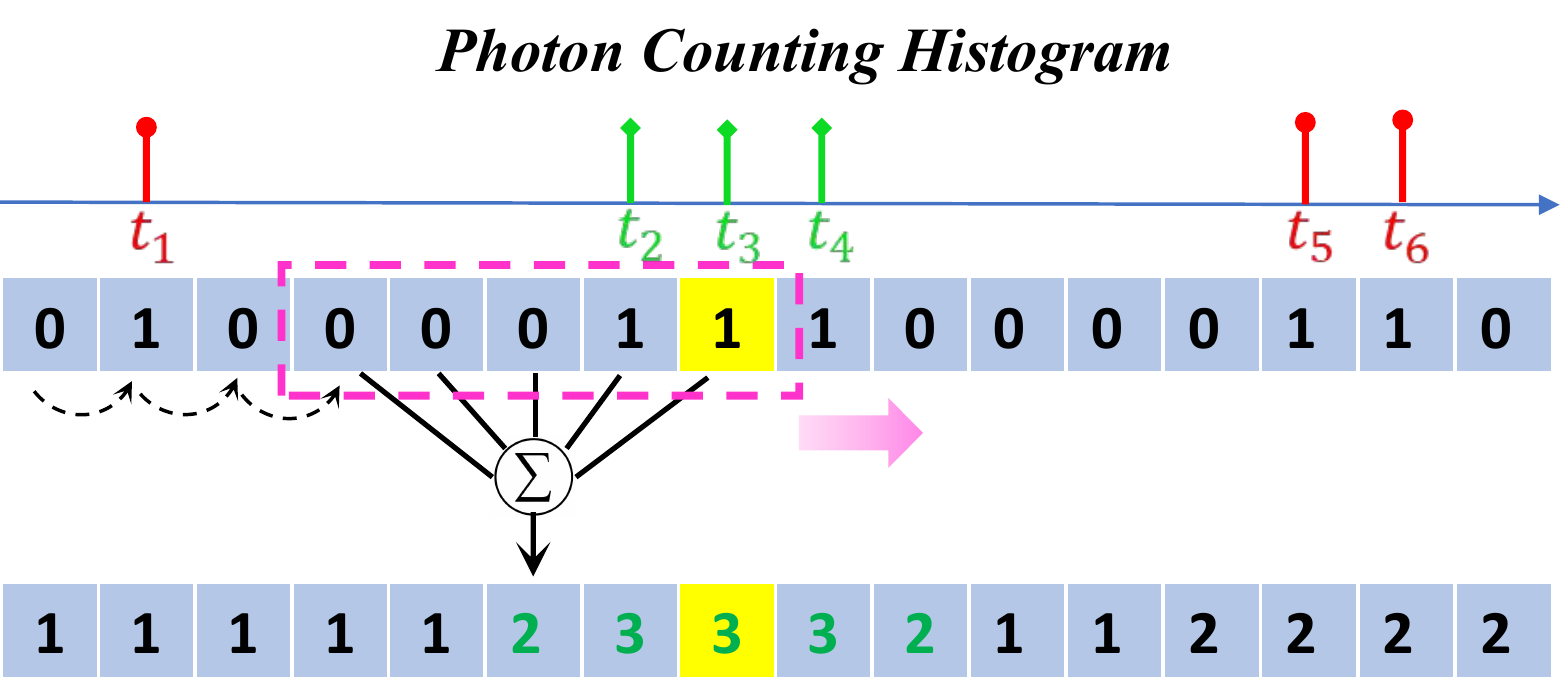}}
	\caption{(a-b) Two examples for the temporal windowing strategy. The raw counting data and the generated clusters are listed as the top and bottom histograms, respectively. The time bin marked in yellow indicates the real depth. The black dashed arrow lines indicate the moving trajectory of the window. Signal and noise photons are marked in green and red, respectively.}
	\label{fig:cluster}
\end{figure}

In contrast to the background noise counts which are uniformly distributed along the time dimension, signal counts tend to cluster around the time bin that corresponds to the true depth. We propose a specific 3D convolutional layer with a kernel of
the dimension $T_{wind} \times 1 \times 1$ to cluster the photon counts. The parameters of this kernel are all initialized as 1 and fixed during the training. For the ($i,j$)-th pixel, the temporal windowing operation is formulated as
\begin{align} 
    h_{ijk} = {\sum_{\ell=-u}^{u}h_{ij(k+\ell)}}, \quad u = \left[\frac{T_{wind}}{2}\right],
\end{align}
where $[\cdot]$ denotes the floor function. As shown in Fig.~\ref{fig:cluster}, each time bin (even without any photon count) can be taken as a candidate for generating the photon cluster. The temporal windowing strategy is not used to directly judge the accurate position of the time
bin that corresponds to the true depth. The goal of temporal windowing is to highlight the signals, such that it becomes much easier for the subsequent modules to find a good filtering threshold than the original input. In other words, candidates for the target time bin are greatly reduced by the temporal windowing strategy. In practice, since most of the signal counts are concentrated in the full width at half maximum (FWHM) of the pulse, we set $T_{wind}$ as close to the FWHM as possible. The temporal windowing strategy in Rapp \textit{et al.} \cite{rapp2017few} calculates a minimum cluster size (a threshold) with a predetermined and complicated formula to filter out the clusters whose counts are below the threshold. In contrast, our temporal windowing is used to enhance the real signal, which facilitates the estimation of soft thresholds to filter out the noise features by the subsequent modules. The thresholds here are generated dynamically and adaptively by the subsequent modules, which may achieve better filtering performance than using thresholds that are determined with a given formula.

\renewcommand{\figurename}{Fig}

\subsubsection{Encoding module}
The encoding module is designed to generate the latent representations from the  photon counting cube. We use the dense dilated fusion strategy \cite{chen2019real} to construct this module. Both normal 3D convolution and dilated 3D convolution are adopted here to extend the receptive field. To reduce the computational and memory consumption, we compress the temporal dimension by four convolutional layers with $2\times1\times1$ stride. Finally, the time dimension of the input is compressed by 4 times and the number of channels is scaled by 32 times.

\subsubsection{PRS block}
\begin{figure}[!t]
	\begin{center}
		\includegraphics[width=0.9\linewidth]{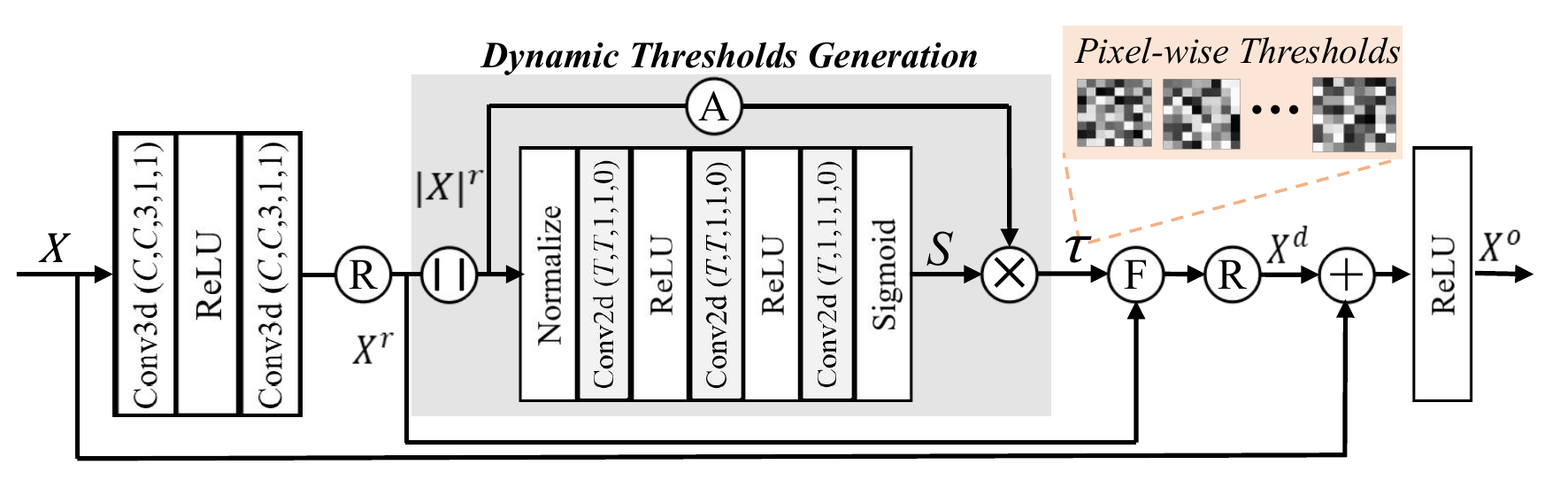}
	\end{center}
	\caption{The pipeline of PRS block. \textcircled{R} is the reshape operation. \textcircled{|\,|} denotes the absolute value function. \textcircled{A} denotes the average operation along the time dimension. \textcircled{x} is the element-wise multiplication. \textcircled{F} is the soft thresholding operation. \textcircled{+} denotes the element-wise summation. For the Convnd($\phi_{1},\phi_{2},\phi_{3},\phi_{4},\phi_{5}$) modules, $\phi_{1}$ is the number of input channels, $\phi_{2}$ is the number of output channels, $\phi_{3}$ is the kernel size, $\phi_{4}$ is the stride and $\phi_{5}$ is the padding. 
	}
	\label{fig:PRS-block}
\end{figure}

In order to improve the feature denoising ability on the high-noise data, the PRS-block uses pixel-wise soft thresholding to boost the SBR on the latent representations. Improved upon the stable threshold generation strategy from \cite{zhao2019deep}, where the same thresholds are shared among all pixels, our strategy is to predict the pixel-wise optimal thresholds for different inputs. As shown in  Fig.~\ref{fig:PRS-block}, the input is denoted by $X \in \mathbb{R}^{B \times C \times T \times M \times N}$, where $B$ is the batch size, $C$ is the number of channels, and $M \times N$ is the input resolution. First, two $3\times3\times3$ convolutional layers followed by a reshape layer are used to obtain the residual input, which is denoted by $X^{r} \in \mathbb{R}^{BC \times T \times M \times N}$, for the pixel-wise soft thresholding operation. Then, $X^{r}$ is fed into the dynamic thresholds generation block, which is composed of two parallel branches. In the first branch, a normalization layer is used at the beginning to improve the training efficiency \cite{ioffe2015batch}. Then three 2D convolutional layers with $1\times1$ kernel size are proposed to calculate the pixel-wise scaling parameter $S \in \mathbb{R}^{BC \times M \times N}$. At the end of this branch, a sigmoid function is used to confine the output within $[0,1]$. In the second branch, we average the absolute values of $X^{r}$ along the time dimension. Then the pixel-wise thresholds $\tau \in \mathbb{R}^{BC \times M \times N}$ are calculated by
\begin{equation}
	\tau_{b,i,j} = S_{b,i,j} \cdot \mathop{\mathrm{average}}\limits_{t}(|X^{r}_{b,t,i,j}|),
\end{equation}
where $b$, $t$, $i$ and $j$ are the corresponding indices in the reshaped tensors. After that, the soft thresholding operation $F(\cdot)$ for each element in $X^{r}$ with its corresponding threshold in $\tau$ is defined by
 \begin{equation}
	F(X^{r}_{b,t,i,j},\tau_{b,i,j}) =
	\begin{cases}
		X^{r}_{b,t,i,j} - \tau_{b,i,j},&\quad X^{r}_{b,t,i,j} > \tau_{b,i,j} \\
		0,&\quad - \tau_{b,i,j} \leq X_{b,i,j} \leq \tau_{b,i,j} \\
		X^{r}_{b,t,i,j} + \tau_{b,i,j},&\quad X^{r}_{b,t,i,j} < -\tau_{b,i,j} 
	\end{cases}.
	\label{Eq:raw_soft}
\end{equation}
Finally, the denoised representation is reshaped as $X^{d} \in \mathbb{R}^{B \times C \times T \times M \times N}$ and added with the input as
\begin{align}
	X^{o} = X + X^{d},
\end{align}
where $X^{o}$ is the denoised output with increased SBR. 

\subsubsection{Decoding module}
The decoding module up-samples the denoised features along the temporal dimension. Four deconvolutional layers with a kernel size of $6\times3\times3$ and a stride of $2\times1\times1$ are adopted here. After that, a $1\times1\times1$ convolutional layer followed by a softmax layer are used to produce the predicted distribution $\hat{p}_{ij}$ of each pixel along the time dimension. The final prediction for the depth $\hat{z}_{ij}$ is generated from $\hat{p}_{ij}$ using a differentiable SoftArgmax \cite{lindell2018single} operation. 

 \subsection{Loss function}
Previous works \cite{lindell2018single, peng2020photon, zang2021non} tried to reconstruct the pulse waveform distribution from sparse and noisy SPAD data at each pixel, as they took the $\mathrm{KL}$ divergence to measure the difference between the output and a target normalized pulse waveform that spans a number of time bins. In fact, the real depth must correspond to a single location within one time bin. Thus, we aim to classify each pixel into its depth category, and therefore the target label is much more centralized than the previous works. Consequently, the produced predictions would be more centralized along the time dimension as well. 

To be more precise, the optimization target is redefined, and a new loss function based on cross-entropy is proposed. Denote
\begin{equation}
	y_{ij} = \left[\frac{2z_{ij}}{\Delta \cdot \mathrm{c}} \right],
\end{equation}
where $z_{ij}$ is the real depth for the $(i,j)$-th pixel and $\Delta$ is the size of time bin. By one-hot encoding, we obtain $P \in \mathbb{R}^{M \times N \times T} $ as the classification label from $\{y_{ij}\}$. Then the cross entropy loss between the output and real label is calculated by
\begin{equation}
	\mathcal{L}_{\mathrm{CE}}(\bm{p_{ij}},\bm{\hat{p}_{ij}}) = -\sum_{k=1}^{T}p_{ijk}\log{\hat{p}_{ijk}}.
\end{equation}
The total loss is given by
\begin{equation}
	\label{Eq:loss}
	\mathcal{L}_{total} = \frac{1}{MN}\sum_{i=1}^{M}\sum_{j=1}^{N}\mathcal{
		L}_{\mathrm{CE}}(\bm{p_{ij}},\bm{\hat{p}_{ij}}) + \lambda \cdot \mathcal{
		L}_{\mathrm{TV}}(\hat{Z}),\quad \lambda>0,
\end{equation}
in which the total variation loss of the predicted depth image is formulated as
\begin{equation}
    \label{Eq:totloss}
	\mathcal{L}_{ \mathrm{TV}}({\hat{Z}})=  \sum_{i=1}^{M}\sum_{j=1}^{N}(|\hat{z}_{i+1, j}-\hat{z}_{i, j}|+|\hat{z}_{i, j+1}-\hat{z}_{i, j}|)
\end{equation}
for spatial smoothing, where $\hat{z}$ is the predicted depth generated from $\hat{p}$. The SoftArgmax function in the following form
\begin{align}
	\hat{z}_{ij} &= \frac{\Delta \cdot \mathrm{c}}{2} \mathrm{SoftArgmax}(\hat{p}_{ij}),\\
	             &= \frac{\Delta \cdot \mathrm{c}}{2}  \sum_{k=1}^{T}k \cdot \hat{p}_{ijk}
	\label{Eq:softargmax}
\end{align}
is employed to approximate the Argmax function. Mathematically, a more centralized distribution enforced by the cross entropy loss will lead to a more accurate approximation to Argmax function, which would further improve the performance of the model. It is clear that a wide $\hat{p}$  (trained by KL) may get an inaccurate result. Therefore, our model classifies each pixel into its depth category to make $\hat{p}$ more centralized than the one trained by fitting the distribution. In other words, although the real signal in a real-world histogram is a distribution, finding the peak of this distribution is the goal of this reconstruction problem. To verify this assumption, we compared the models trained with KL and CE in our ablation study in Section \ref{Sec:ablation}. The model trained by CE achieved better performance than the one trained by KL.

\section{Experiment}
In this section, we first introduce the implementation details and evaluation metrics. After that, the comprehensive evaluation of the proposed model in comparison with the state-of-the-art approaches on both simulated and real-world data is presented. Results of the ablation study are given at the the end of this section.

\subsection{Implementation details}
\paragraph{Data simulation.} The simulated data of photon counting cubes are sampled using the observation model, in which the ground-truth depth images and reflectivity images are adopted from the NYU v2 dataset \cite{silberman2012indoor}. NYU v2 consists of a large number of video sequences captured with Microsoft Kinect sensor from various indoor scenarios. In line with the previous works, the total number of time bins is set as $1024$ and the size of time bins is 80$ps$. The FWHM of the pulsed light is 400$ps$. All the images are reshaped into a resolution of $64 \times 64$. As a result, the size of the simulated counting cube is $64\times 64 \times 1024$. 13000 and 3000 samples are generated for training and validating, respectively.

\paragraph{Train Setting.}The deep learning models illustrated in this section are implemented using the deep learning library PyTorch on NVIDIA 1080 GPU. Adam optimizer \cite{kingma2015adam} is adopted with an initial learning rate of $1 e^{-3}$ and a decay rate of $0.6$. We set the $\lambda$ in Eq. (\ref{Eq:loss}) as $1 e^{-6}$. In particular, $32 \times 32 \times 1024$ patches are cropped randomly from the simulated cubes as the training inputs for deep learning models. The batch size is set as 4. We train the PRS-Net on data across different SBR conditions denoted by (10:2, 5:2, 2:2, 10:10, 5:10, 2:10, 10:50, 5:50, 2:50, 3:100, 2:100, 1:100).

\paragraph{Test Setting.} All the approaches, including the deep learning and statistical learning models are firstly evaluated on the Middlebury stereo dataset \cite{scharstein2007learning}, which consists of eight scenes including Art, Books, Bowling, Dolls, Laundry, Moebius, Plastic and Reindeer. We resize the images to $576 \times 640$ (for Bowling \& Plastic scenes) and $576 \times 704$ (for other scenes) pixels resolution. Then, experiments are conducted on real-world data with $256 \times 256 \times 1536$ pixels resolution measured by \cite{lindell2018single}, which is resized into $256 \times 256 \times 1024$. For deep learning models, the evaluations are conducted in a patch-by-patch manner, where each input is cropped into $128\times128 \times 1024$ patches with a stride of $64$. After that, the recovered patches are assembled together to produce a full resolution output.

\renewcommand{\tablename}{Table}
\begin{table}[!ht]
	\centering
	\caption{Performance comparisons on the simulated dataset. The results of two metrics are reported as the average over the eight scenes from Middlebury dataset. The best results for each column are highlighted in bold. The statistic learning models are run on CPU (48GB) while deep learning models are run on GPU (8GB). The computational costs during the test are summarized as well. }
	\label{tab_1}
	\small
	\renewcommand\arraystretch{1}
	\begin{tabular}{p{2.2cm}<{\centering}|p{1.3cm}<{\centering}p{1.3cm}<{\centering}p{1.3cm}<{\centering}p{1.3cm}<{\centering}p{1.3cm}<{\centering}p{1.5cm}<{\centering}}
		\toprule
		\multicolumn{7}{c}{Signal photons : 2 \quad Background photons : 10 \quad SBR : 0.2}                                   \\
		\toprule
		&Error               & \multicolumn{3}{c}{Accuracy with $\delta$}                  &\multicolumn{2}{c}{Computational cost}  \\
		\cmidrule(r){1-2} \cmidrule(r){3-5} \cmidrule(r){6-7}
		Methods       &RMSE(m)           & $\delta = 1.01$           & $\delta = 1.02$           &$\delta = 1.03$         & Time(s) & Memory(GB) \\
		\midrule
		LM filter     & 4.7143          & 41.73\%          & 52.89\%          & 54.80\%          &\textbf{11.87}       &3.10            \\
		Shin \textit{et al.}    & 4.3205          & 0.00\%           & 0.00\%           & 0.00\%           &45.70      &3.60         \\
		Rapp \textit{et al.}    & 0.0574          & 91.84\%          & 97.16\%          & 97.88\%         &182.17      &6.30           \\
		U-Net & 0.0416          & 19.85\%          & 68.54\%          & 95.78\%          &1475.06  &7.35            \\
		U-Net++       & 0.0308          & 89.04\%          & 98.90\%          & 99.35\%          &1085.83  &7.28            \\
		ADAG-U-Net++  & 0.0248          & 92.52\%          & 99.09\%          & 99.45\%          &3367.24  &4.65            \\
		Non-local & 0.0155          & 86.84\%          & 99.17\%          & \textbf{99.61\%}     &139.14   &3.59            \\
		PRS-Net           & \textbf{0.0119} & \textbf{96.99\%} & \textbf{99.46\%} & 99.60\%          &95.44     &\textbf{2.68}            \\
		\midrule
		\multicolumn{7}{c}{Signal photons : 2 \quad  Background photons : 50 \quad SBR : 0.04}                                  \\
		\midrule
		LM filter     & 5.7796          & 28.07\%          & 34.79\%          & 35.85\%          &\textbf{12.58}       &3.70            \\
		Shin \textit{et al.}   & 5.4726          & 0.00\%           & 0.00\%           & 0.00\%           &62.21       &7.50            \\
		Rapp \textit{et al.}    & 0.0994          & 89.46\%          & 95.58\%          & 97.38\%         &274.06      &11.80          \\
		U-Net & 0.0663          & 28.74\%          & 81.40\%          & 93.79\%          &1475.06  &7.35            \\
		U-Net++       & 0.0357          & 88.92\%          & 98.31\%           & 98.92\%          &1085.83  &7.28            \\
		ADAG-U-Net++  & 0.0310          & 90.07\%          & 98.63\%          & 99.15\%          &3367.24  &4.65            \\
		Non-local & 0.0172          & 86.43\%          & 98.64\%          & 99.31\%              &139.14   &3.59            \\
		PRS-Net           & \textbf{0.0133} & \textbf{96.12\%} & \textbf{99.20\%} & \textbf{99.55\%} &95.44    &\textbf{2.68}            \\
		\midrule
		\multicolumn{7}{c}{Signal photons : 2 \quad  Background photons : 100 \quad SBR : 0.02}                                 \\
		\midrule
		LM filter     & 6.2041          & 20.68\%          & 25.49\%          & 26.31\%          &\textbf{13.28}       &3.90            \\
		Shin \textit{et al.}   & 5.622           & 0.00\%           & 0.00\%           & 0.00\%           &95.11       &9.30            \\
		Rapp \textit{et al.}    & 0.1779          & 87.4\%           & 94.19\%          & 96.36\%          &658.18     &18.50            \\
		U-Net & 0.079           & 34.06\%          & 84.16\%          & 94.58\%          &1475.06  &7.35          \\
		U-Net++       & 0.0478          & 90.01\%          & 97.44\%          & 98.25\%          &1085.83  &7.28           \\
		ADAG-U-Net++  & 0.0425          & 88.24\%          & 97.35\%          & 98.28\%          &3367.24  &4.65            \\
		Non-local & 0.0255          & 83.56\%          & 98.09\%          & 99.04\%          &139.14   &3.59           \\
		PRS-Net           & \textbf{0.0152} & \textbf{95.09\%} & \textbf{98.93\%} & \textbf{99.39\%} &95.44    &\textbf{2.68}            \\
		\midrule
		\multicolumn{7}{c}{Signal photons : 1 \quad  Background photons : 100 \quad SBR : 0.01}                                 \\
		\midrule
		LM filter     & 6.8807          & 6.75\%           & 8.89\%           & 9.46\%        &\textbf{13.12}       &3.80            \\
		Shin \textit{et al.}   & 5.6959          & 0.00\%           & 0.00\%           & 0.00\%       &80.37        &8.70            \\
		Rapp \textit{et al.}    & 0.6658          & 75.21\%          & 85.23\%          & 89.75\%          &972.88  &25.20            \\
		U-Net & 0.4085          & 33.32\%          & 73.41\%          & 85.25\%          &1475.06  &7.35           \\
		U-Net++       & 0.2811          & 77.24\%          & 85.70\%          & 88.43\%          &1085.83  &7.28           \\
		ADAG-U-Net++  & 0.1899          & 79.37\%          & 91.26\%          & 93.29\%          &3367.24  &4.65           \\
		Non-local& 0.0706          & 71.04\%          & 92.36\%          & 95.54\%          &139.14   &3.59            \\
		PRS-Net          & \textbf{0.0424} & \textbf{87.51\%} & \textbf{96.54\%} & \textbf{98.01\%} &95.44    &\textbf{2.68}            \\
		\bottomrule
	\end{tabular}
\end{table}

\subsection{Evaluation metrics}
To compare the performances of different approaches, two commonly-used metrics \cite{zang2021non,peng2020photon,lindell2018single} are adopted, namely, the root mean-square error (RMSE) which is formulated as 
\begin{equation}
	\mathrm{RMSE} = \sqrt[]{ \frac{1}{MN}\sum_{i=1}^{M}\sum_{j=1}^{N} {(z_{ij} - \hat{z}_{ij})}^{2} },
\end{equation}
and the accuracy at a given threshold $\delta$ which is formulated as
\begin{equation}
	\mathrm{percentage \,\, of \,\,} \hat{Z} \mathrm{\,\, s.t. \,\,}
	\frac{1}{MN}\sum_{i=1}^{M}\sum_{j=1}^{N} {\mathrm{max}(\frac{\hat{z}_{ij}}{z_{ij}}, \frac{z_{ij}}{\hat{z}_{ij}})<\delta}
	\label{Eq:accuracy}
\end{equation}

\subsection{Approaches for comparison} Approaches for comparison are:
\begin{itemize}
    \item{\textbf{LM filetr} \cite{bar1969communication}: Conventional maximum likelihood estimation.}
    \item{\textbf{Shin \textit{et al.}} \cite{shin2015photon}: The maximum likelihood estimation with a total variance regularization term.}
    \item{\textbf{Rapp \textit{et al.}} \cite{rapp2017few}: The maximum likelihood estimation combined with a total variance regularization term, a cluster searching algorithm and a superpixel formation strategy.}
    \item{\textbf{U-Net}\cite{lindell2018single}: The deep learning model based on U-Net \cite{ronneberger2015u}. To make a fair comparison, we adopt the version without the intensity fusion strategy.}
    \item{\textbf{Non-local} \cite{peng2020photon}: The deep learning model based on non-local blocks \cite{wang2018non} and deep boosting backbones \cite{chen2019real}.}
    \item{\textbf{U-Net++} \cite{zang2021non}: The deep learning model based on U-Net++ \cite{zhou2019unet++}.}
    \item{\textbf{ADAG-U-Net++}\cite{zhao2022photon}: The deep learning model based on U-Net++ with an Attention-Directed Attention Gate (ADAG) module to enhance the edges' reconstruction.}
\end{itemize}

\begin{figure}[t]
\begin{center}
\subfloat[]{\includegraphics[width=0.5\linewidth]{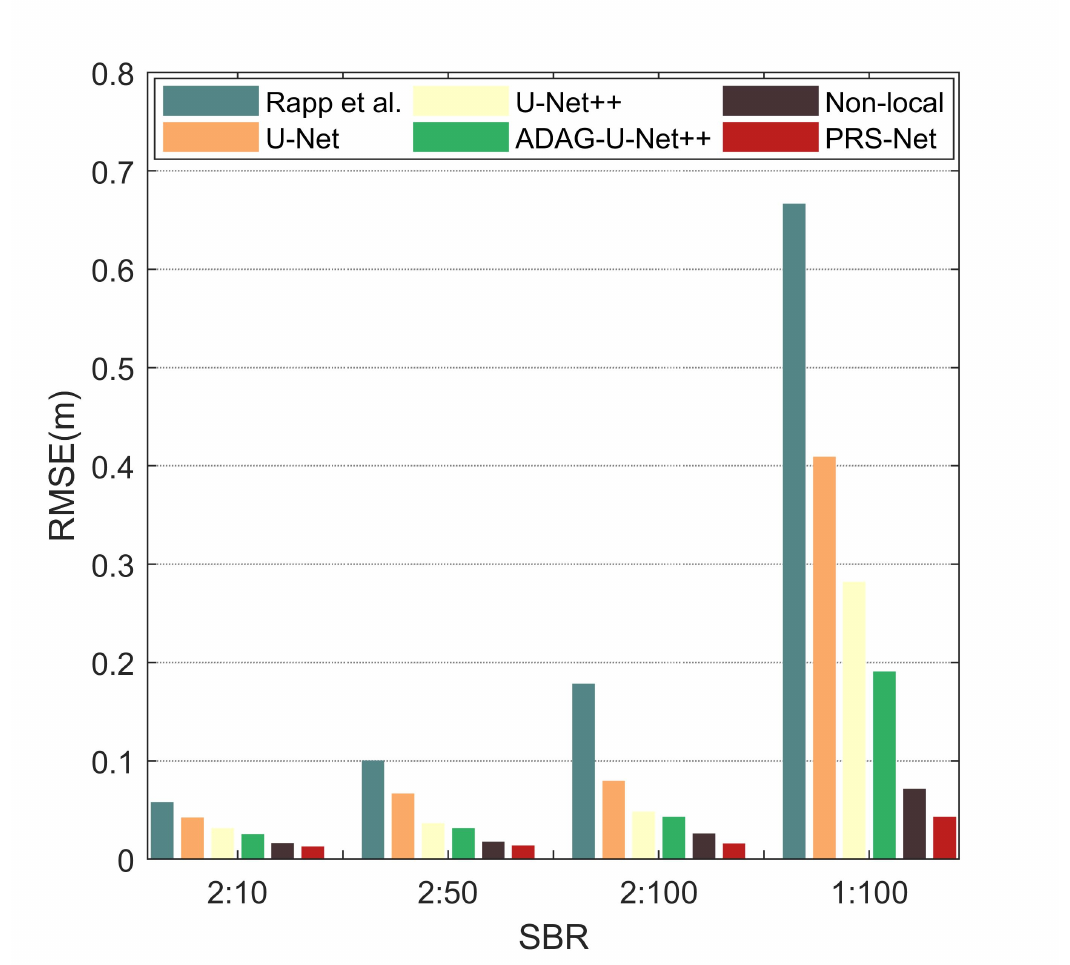}}
\subfloat[]{\includegraphics[width=0.5\linewidth]{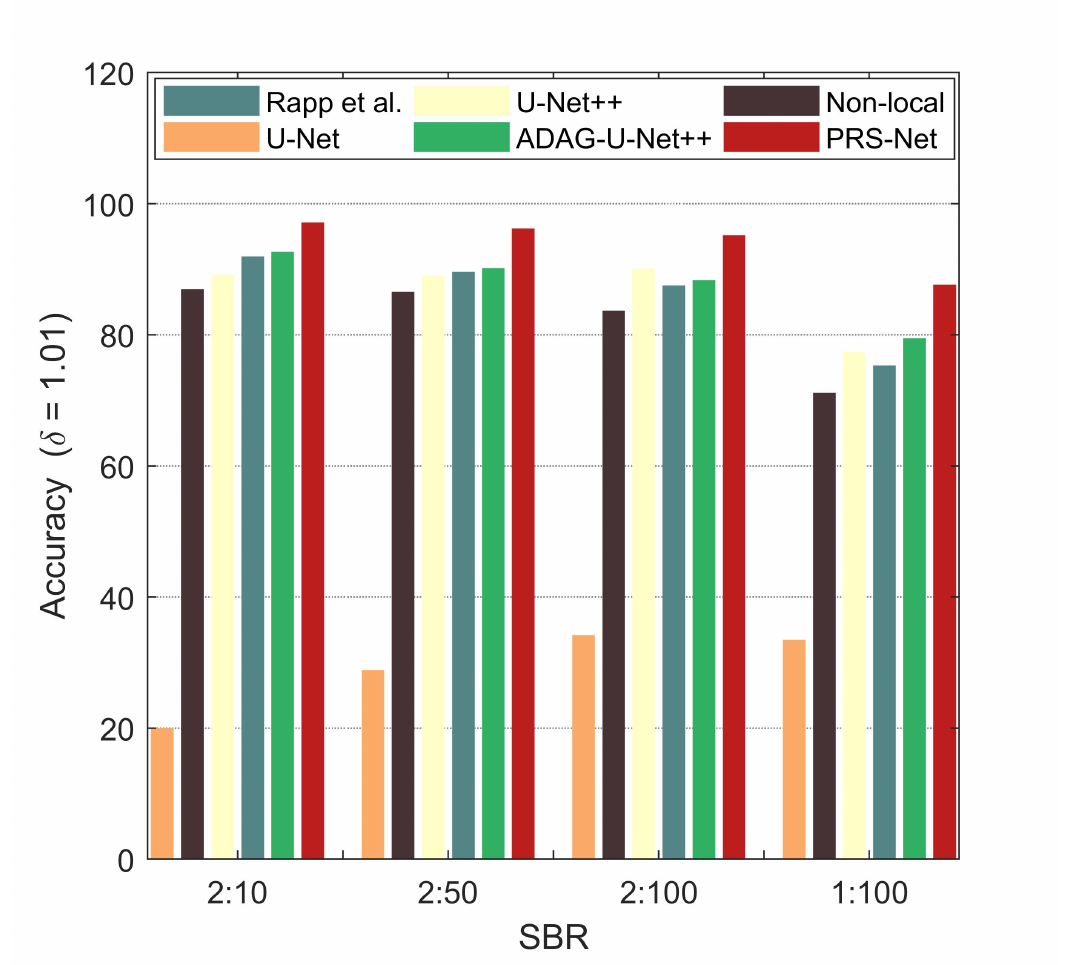}}
\end{center}
\vspace{-6mm}
\caption{ (a) RMSE values obtained with four SBRs. (b) Accuracies obtained with $\delta = 1.01$ and four SBRs.}
\label{fig:trend}
\end{figure}

\subsection{Evaluation on simulated data}
The test results on the Middlebury dataset are generalized in Table \ref{tab_1}. Under two typical SBR levels (i.e., 2:10 and 2:50) and two extremely low SBR levels (i.e., 2:100 and 1:100), the PRS-Net achieves the best RMSE values among all the models of all cases. Considering the accuracy metrics, except for the case with SBR of 2:10 and $\delta$ of 1.03 where Non-local net slightly outperforms PRS-Net, PRS-Net beats the other models by a large margin. Specifically, PRS-Net have achieved an accuracy of more than 95\% with $\delta=1.01$ when SBRs are 2:10, 2:50 and 2:100. Even in the extreme case that SBR is 1:100, PRS-Net still achieves an accuracy of 87.51\%, which is significantly better than other models. The comparison results in terms of running time and maximum GPU memory are also displayed in Table \ref{tab_1}. The statistical learning algorithms are run on 48G RAM and Intel i5-9400 CPU, which uses 6 parallel kernels. It should be noted that during the test, statistical approaches use the original high-resolution data instead of patches as the input since CPU memory is large enough. It should be noted that since the $128\times128\times1024$ resolution always used up the memory of our GPU, we tested ADAG-U-Net++ with a smaller resolution of $64\times64\times1024$. It can be seen that PRS-Net is the fastest among all the deep learning models. Besides, PRS-Net requires the least amount of memory among all the models, including both the statistical and deep learning ones.

\begin{figure}[!htp]
\begin{center}
\subfloat[]{\includegraphics[width=\linewidth]{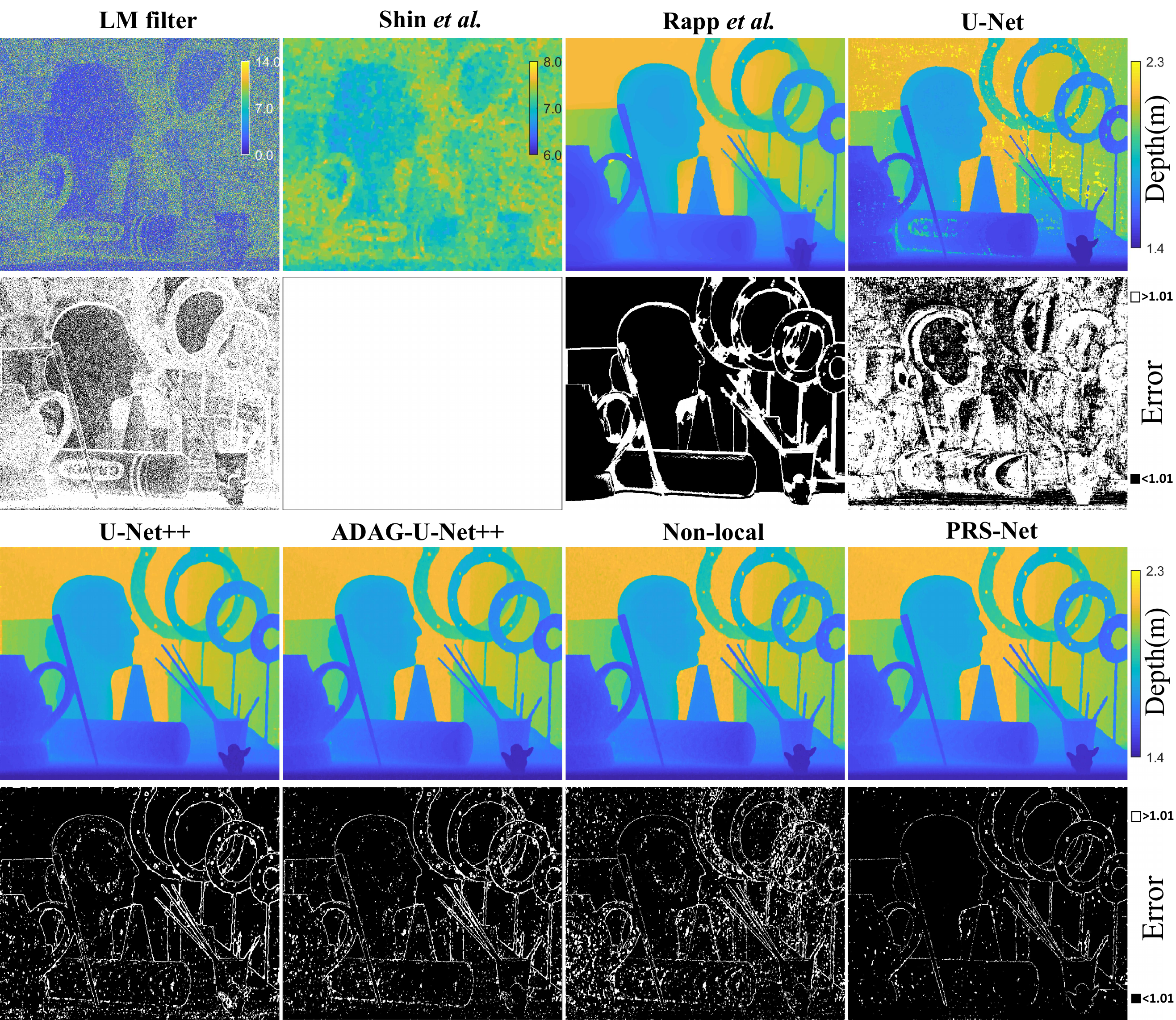}}\\
\vspace{-2mm}
\subfloat[]{\includegraphics[width=\linewidth]{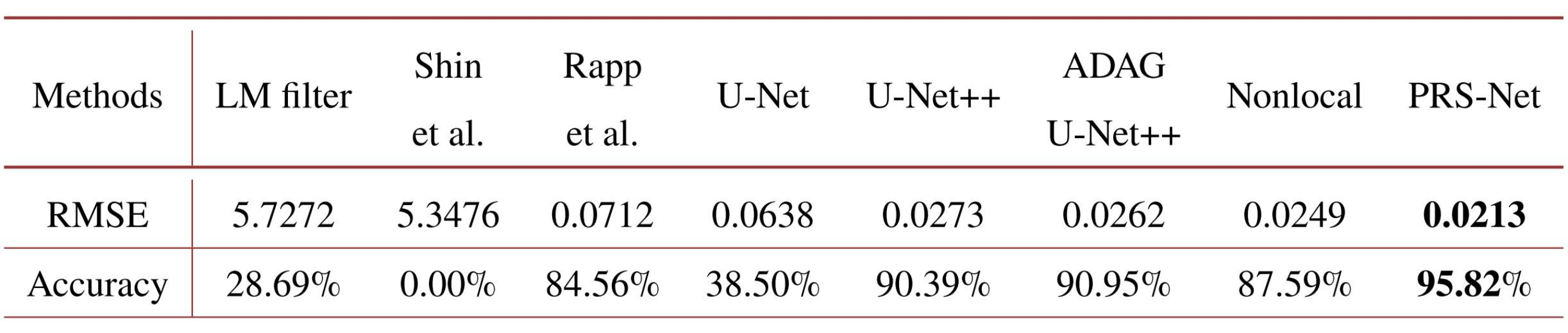}}
\end{center}
\vspace{-5mm}
\caption{
Comparison results of different approaches in the Art scene with SBR of 2:50. (a) The pixels of RGB images are colored using the predicted depth values with the colorbars. The corresponding monochrome images are the visualizations of the error maps. The white pixels are ones whose predictions fail to satisfy Eq. (\ref{Eq:accuracy}) for $\delta = 1.01$. Alternatively, the differences between the predicted and real depths of the black pixels are below 1\%. Note that the depth images created by LM filter and Shin \textit{et al.} are colored with different scales due to high variance of the predictions. (b) The RMSE and Accuracy (for $\delta=1.01$) results calculated on the Art Scene with SBR of 2:50. The best results are highlighted in bold.}
\label{fig:sim_fig_1}
\end{figure}

More importantly, as compared with other competitors, PRS-Net is robust against noise. We depict the curves of RMSE values and accuracies with $\delta = 1.01$ by varying SBR in Fig.~\ref{fig:trend} (a) and (b), respectively. The performances of other models degrade significantly as noise level increases, while the PRS-Net maintains stable performance, despite a slight drop in the extreme case of SBR$=0.01$. In this case, as compared with the best competitor, the RMSE value of PRS-Net is reduced by $40\%$ and the accuracy is improved by $8.5\%$.

\begin{figure}[!htp]
\begin{center}
\subfloat[]{\includegraphics[width=\linewidth]{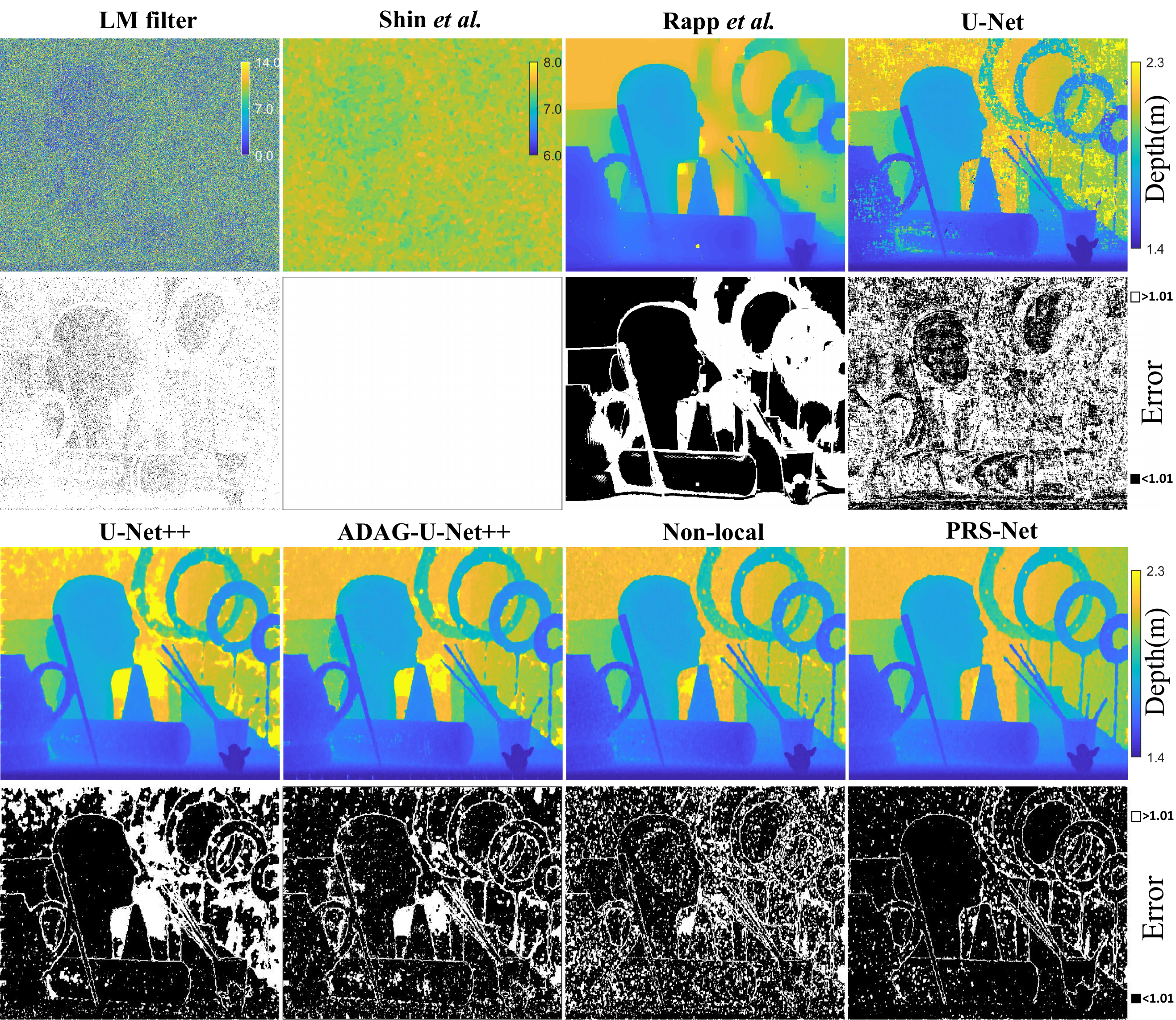}}\\
\vspace{-2mm}
\subfloat[]{\includegraphics[width=\linewidth]{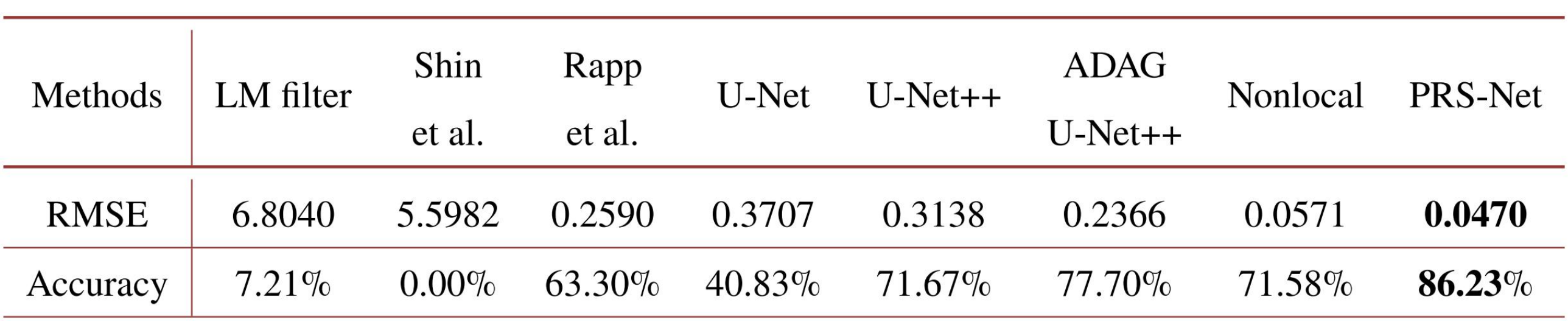}}
\end{center}
\vspace{-5mm}
\caption{
Comparison results of different approaches in the Art scene with SBR of 1:100. (a) The pixels of RGB images are colored using the predicted depth values with the colorbars. The corresponding monochrome images are the visualizations of the error maps. The white pixels are ones whose predictions fail to satisfy Eq. (\ref{Eq:accuracy}) for $\delta = 1.01$. (b) The RMSE and Accuracy (for $\delta=1.01$) results calculated on the Art Scene with the same SBR of 1:100. The best results are highlighted in bold.}
\label{fig:sim_fig_2}
\end{figure}

Examples of recovered depth images from Art scene and visualizations of the corresponding error maps with 2:50 and 1:100 SBRs are shown in Fig.~\ref{fig:sim_fig_1} and Fig.~\ref{fig:sim_fig_2}, respectively. It is clear that PRS-Net achieves the best imaging quality under both SBRs. According to the error maps in which black pixel indicates a less than 1\% difference between the real depth and prediction, the number of abnormal pixels in the depth images of PRS-Net is the least among all competitors, especially in the extreme case of SBR=0.01. In particular, the accuracy of PRS-Net with $\delta=1.01$ has reached 95.82\% for SBR of 2:50, while the accuracy of the second-best model is 90.95\%. Although the accuracy of PRS-Net has fallen below 90\% for SBR of 1:100, it is still the only approach that maintains an accuracy beyond 85\%.

In order to measure and eliminate the randomness induced by Poisson sampling process, we choose the Art and Bowling scenes from Middlebury stereo dataset as the representatives of complex and simple scenes, and simulate 10 sets of random data with different SBRs to calculate the average performance. More specifically, 10 trials are repeated to calculate the average RMSE values with 10 sets of randomly generated data based on Poisson sampling. As shown in Table \ref{tab_AB},  PRS-Net achieves superior performance which is consistent with single test. Moreover, PRS-Net basically has the smallest standard deviations when compared with other models, which suggests a remarkable robustness of performance for PRS-Net.

\begin{table}[!htp]
\centering
\caption{The mean and standard deviation of RMSE results in Art and Bowling scenes by repeating ten trails, for different SBRs. For each column, the best results are highlighted in bold. }
\label{tab_AB}
\small
\begin{tabular}{c|c|c|c|c}
\toprule
Methods         & 2:10          & 2:50          & 2:100         & 1:100         \\ \toprule
LM filter   & 4.7528\,$\pm$\,0.0061 & 5.8272\,$\pm$\,0.0059 & 6.2455\,$\pm$\,0.0053 & 6.9316\,$\pm$\,0.0059 \\ \hline
Shin \textit{et al.} & 4.3796\,$\pm$\,0.0043 & 5.5348\,$\pm$\,0.0020 & 5.6837\,$\pm$\,0.0009 & 5.7593\,$\pm$\,0.0014 \\ \hline
Rapp \textit{et al.} & 0.0442\,$\pm$\,0.0129 & 0.0498\,$\pm$\,0.0071 & 0.0995\,$\pm$\,0.0593 & 0.4369\,$\pm$\,0.1116 \\ \hline
U-Net  & 0.0411\,$\pm$\,0.0003 & 0.0618\,$\pm$\,0.0010 & 0.0692\,$\pm$\,0.0046 & 0.4110\,$\pm$\,0.0398 \\ \hline
U-Net++  & 0.0208\,$\pm$\,0.0002 & 0.0258\,$\pm$\,0.0010 & 0.0315\,$\pm$\,0.0023 & 0.2291\,$\pm$\,0.0305 \\ \hline
ADAG-U-Net++     & 0.0189\,$\pm$\,0.0002 & 0.0245\,$\pm$\,0.0004 & 0.0292\,$\pm$\,0.0007 & 0.1408\,$\pm$\,0.0328 \\ \hline
Non-local   & 0.0194\,$\pm$\,0.0002 & 0.0223\,$\pm$\,0.0003 & 0.0248\,$\pm$\,0.0004 & 0.0465\,$\pm$\,0.0072 \\ \hline
PRS-Net        & \textbf{0.0162\,$\pm$\,0.0003} & \textbf{0.0191\,$\pm$\,0.0003} & \textbf{0.0220\,$\pm$\,0.0005} & \textbf{0.0351\,$\pm$\,0.0021} \\ \hline
\end{tabular}
\end{table}

\begin{figure}[!htp]
\begin{center}
\includegraphics[width=0.9\linewidth]{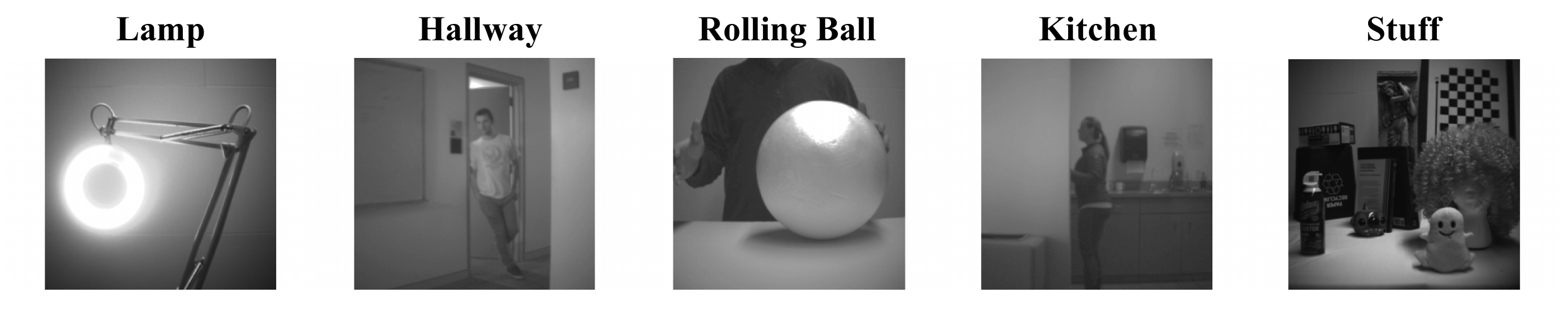}
\includegraphics[width=\linewidth]{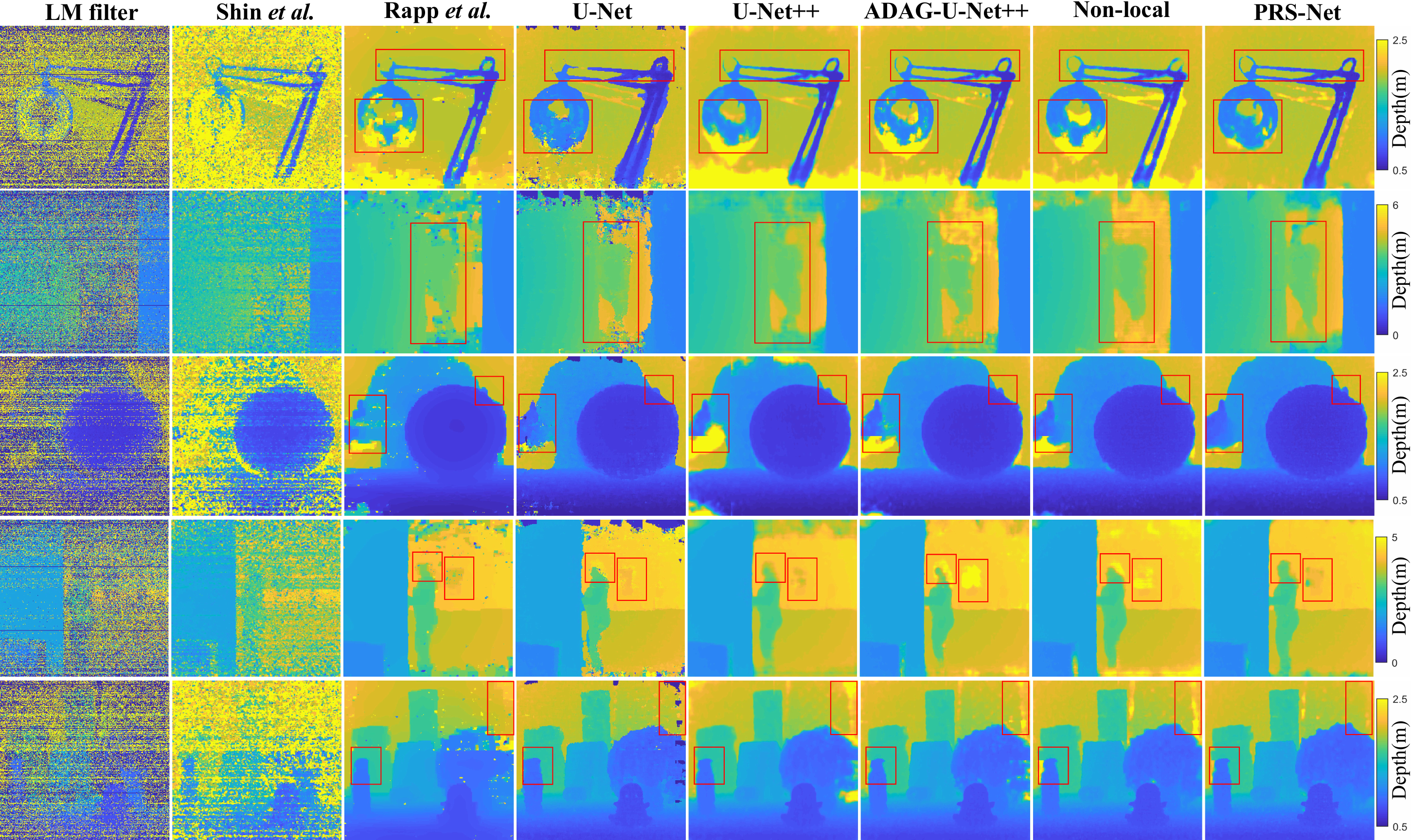}
\end{center}
\caption{
The reconstructed depth images on five real-world scenarios. The first row presents the intensity images captured by a regular camera.}
\label{fig:real_fig}
\end{figure}

\subsection{Real-world evaluation}

We conduct experiments on real-world data captured by indoor single-photon imaging prototype \cite{lindell2018single}, where the photon counting histogram for each pixel consists of 1536 time bins with a bin size of 26$ps$. In the real-world data, photon counts are separately distributed in several consecutive bins. To ease processing, we sum the photon counts of every neighboring pair of bins along the time dimension to get 768 bins. By this way, the number of time bins is compressed to 1024, with the remaining bins padded by zeros. Thus the bin size has doubled to 52$ps$.

Figure \ref{fig:real_fig} shows the performance comparisons in five real-world scenarios, including Lamp, Hallway, Rolling ball, Kitchen and Stuff. In Lamp scene, PRS-Net produces the cleanest image, with the most precise details around the bulb and iron arm. In Hallway scene, PRS-Net is the only approach that has successfully restored the complete human body. In Ball scene, PRS-Net produces the least amount of noise near the left palm, and the clearest thumb on the right. In Kitchen scene, PRS-Net recovers the clearest outline of the human head, such that even the nose can be distinguished. In addition, the black box on the wall has been successfully recovered as well, which is the closest to the intensity image among all the models. In Stuff scene, PRS-Net has produced the fewest abnormalities.

\subsection{Distribution of predictions with pixel-wise approach}

The improvement brought by PRS-Net can be understood by calculating the averaged variance of the probability distribution of predictions along the time dimension by
\begin{align} 
    \label{Eq:energy}
    V = \frac{1}{MN} \sum_{i=1}^{M} \sum_{j=1}^{N} \left(\sum_{k=1}^{T}\hat{p}_{ijk}(k - \bar{k}_{ij})^{2} \right),
\end{align}
where
\begin{equation}
    \bar{k}_{ij} = \sum_{k=1}^{T}\hat{p}_{ijk} \cdot k.
\end{equation}
Variance defined in Eq. (\ref{Eq:energy}) measures the dispersion of the probability distribution over all the pixels, which indicates how confident the model is about its prediction. In PRS-Net, the cross-entropy loss for pixel-wise classification makes the probability distribution of predictions more centralized at one time bin, which leads to a more precise approximation to Argmax function with the SoftArgmax operation in Eq. (\ref{Eq:softargmax}). The comparison results on the averaged variances using Eq. (\ref{Eq:energy}) in the Art scene are summarized in Table \ref{tab_2}. It can be seen that the averaged variances of predictions of PRS-Net is much lower than other deep learning models for all values of SBRs, which means that the predictions of PRS-Net are more  centralized. 

\begin{figure}[!htp]
\begin{center}
\includegraphics[width=\linewidth]{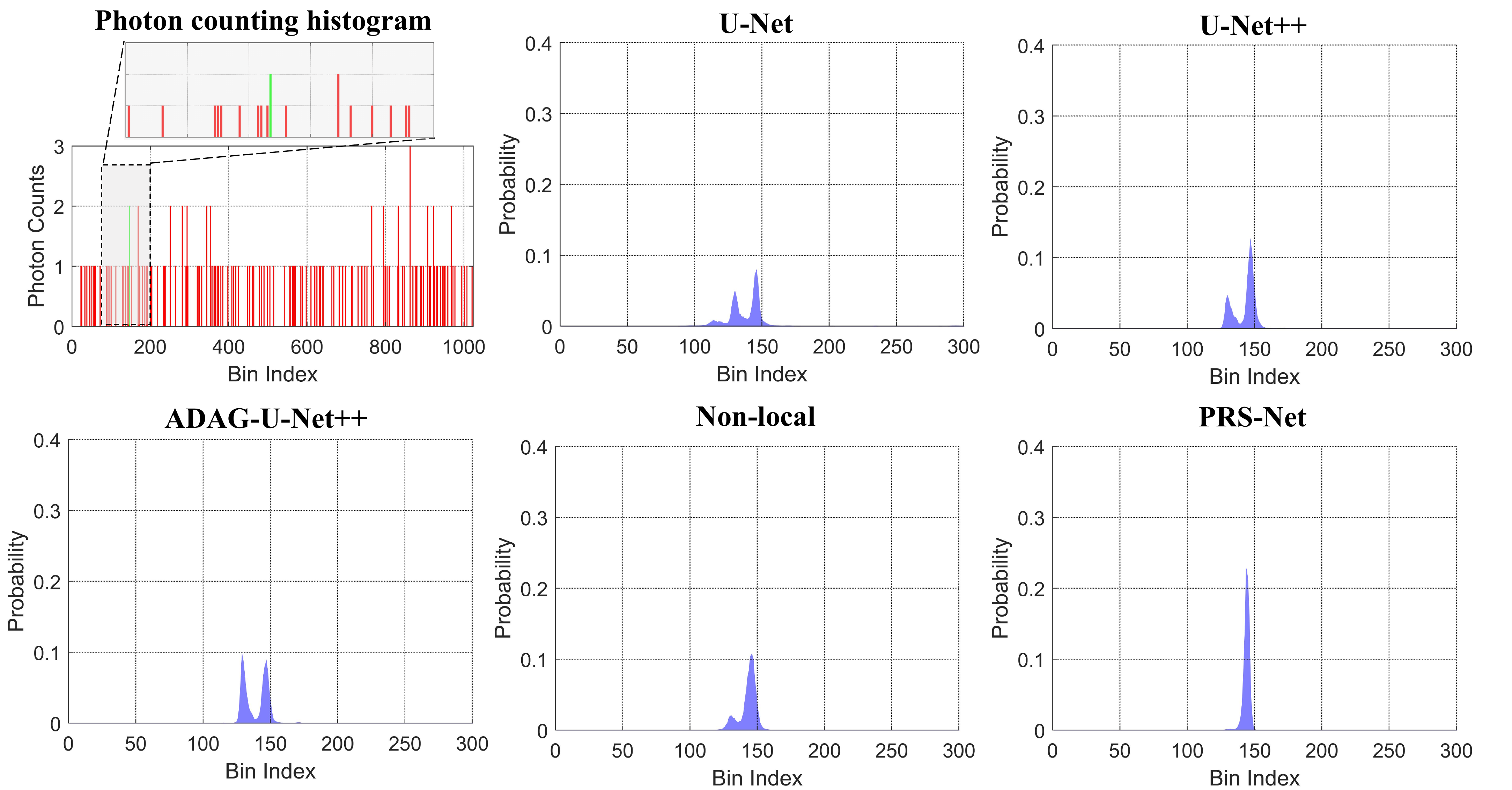}
\end{center}
\caption{Visualization of the distributed predictions for five deep learning models including PRS-Net with SBR=0.01. The data are sampled at the (187,157)-th pixel from Art scene. Signal and noise photon counts in the raw photon counting histogram are marked in green and red, respectively.}
\label{fig:energy}
\end{figure}

In particular, we randomly select a pixel from the Art scene and visualize the output distributions of predictions produced by different models in Fig.~\ref{fig:energy}. It is clear that the width of the probability distribution produced by PRS-Net is the smallest, and the peak of distribution is the highest among all competitors. 

\begin{table}[ht]
\centering
\caption{Averaged variances of the output probability distributions for different deep learning models in Art scene. The average is taken over all pixels. The best result for each column is highlighted in bold.}
\label{tab_2}
\small
\begin{tabular}{c|c|c|c|c}
\toprule
Methods       & 2:10  & 2:50  & 2:100 & 1:100  \\ \toprule
U-Net         & 19.46 & 21.88 & 29.72 & 79.46  \\ \hline
U-Net++       & 17.34 & 17.41 & 17.63 & 23.70 \\ \hline
ADAG-U-Net++  & 17.46 & 17.49 & 17.81 & 20.71  \\ \hline
Non-local     & 17.39 & 17.74 & 17.90 & 18.90  \\ \hline
PRS-Net       & \textbf{7.57}  & \textbf{8.15}  & \textbf{8.81}  & \textbf{12.33}  \\ \hline
\end{tabular}
\end{table}

\subsection{Ablation study}
\label{Sec:ablation}
\renewcommand{\tablename}{\textcolor[rgb]{0.00,0.00,0.00}{Table}}
\begin{table}[ht]
\centering
\caption{Ablation study results using RMSE metric on the simulated dataset. PRS-Net is tested in different combination of components, where A stands for the temporal window strategy, B stands for the soft-thresholding operation in PRS-block, C and D stand for training with pulse waveform reconstruction loss and pixel-wise classification loss, respectively. $\checkmark$ indicates that this component is adopted in the network. The best results are highlighted in bold and comparable results are underlined. }
\label{tab_ablation}
\small
\begin{tabular}{p{15 pt}p{15 pt}p{15 pt}p{15 pt}|cccc}
\toprule
\multicolumn{4}{c|}{Experimental Variables}                                            & \multicolumn{4}{c}{RMSE(m) with Different SBRs}                                                                                                      \\
\toprule
A           & B           &C            & D           & 2:10           & 2:50         & 2:100           & 1:100          \\
\toprule
\checkmark  & \checkmark  & \checkmark  &             & 0.0184         & 0.0212       & 0.0281           & 0.0924                     \\
\hline
\checkmark  &             &             & \checkmark  & \underline{0.0117}   & \underline{0.0132}      & 0.0247      & 0.1326                     \\
\hline
             &\checkmark &             & \checkmark  & \underline{0.0119} & \underline {0.0132}    & 0.0155       & 0.0491 \\
\hline
\checkmark  & \checkmark  &            & \checkmark   & \underline{0.0119} & \underline{0.0133}   & \textbf{0.0152}     & \textbf{0.0424}   \\
\hline
\end{tabular}
\end{table}

\begin{figure}[!ht]
\begin{center}
\includegraphics[width=\linewidth]{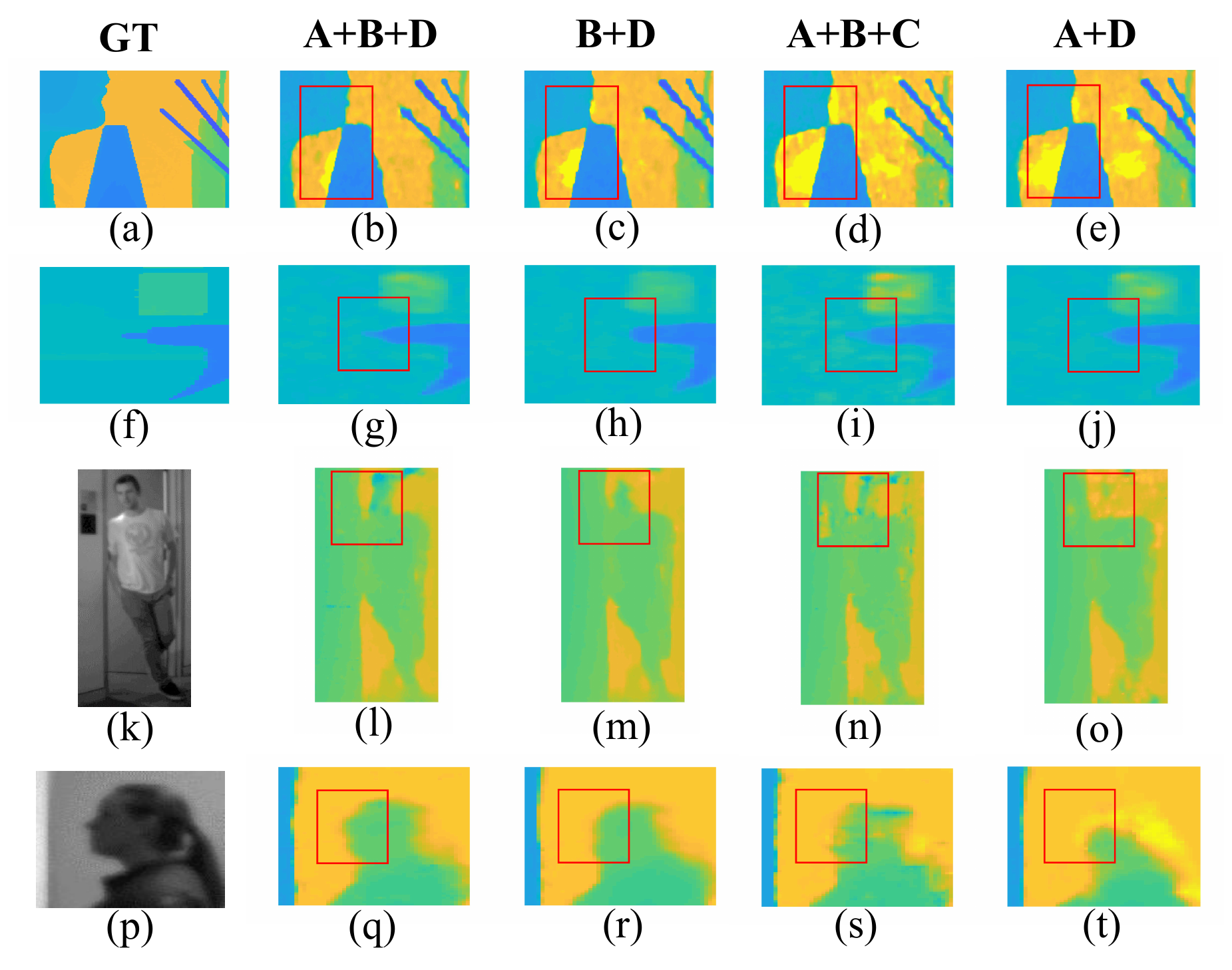}
\end{center}
\vspace{-5mm}
\caption{The definitions of A, B, C and D are the same as in Table.~\ref{tab_ablation}. The top two rows are the reconstructed depth images on simulated data with SBR=0.01 and the bottom two rows are on real-world data. (a$\thicksim$e) are cropped from Art. (f$\thicksim$j) are cropped from Laundry. (k$\thicksim$o) are cropped from Hallway. (p$\thicksim$t) are cropped from Kitchen. Note that we use depth images as the ground-truths in (a) and (f) while use intensity images in (k) and (p).}
\label{fig:ablation}
\end{figure}

To verify the effectiveness of soft-thresholding operation in PRS-block, pixel-wise classification loss and temporal window strategy, we conduct ablation tests by removing the relevant components from the architecture as depicted in Fig.~\ref{fig:overview}. By comparing the results of the second and last row in Table \ref{tab_ablation}, it can be seen that the network with and without the soft-thresholding branch in PRS-block achieve comparable performances if SBR is 2:10 or 2:50. However, under the high-noise setting, i.e., SBR of 2:100 or 1:100, the complete model with soft-thresholding outperforms the model without soft-thresholding branch by a large margin. By comparing the results of the first and last row, we see that the pixel-wise classification loss greatly improves the performance of the model over all SBRs, which demonstrates the advantage of using pixel-wise approach. Besides, the results of the third and last row show that the temporal window strategy may slightly decrease the RMSE. 

We also show the ablation test results on both the simulated data and the real-world data in Fig.~\ref{fig:ablation}. The results are consistent with the statistics in Table~\ref{tab_ablation}. The differences between CE and KL losses are shown by comparing the columns A+B+D and A+B+C. The differences between with and without the soft-thresholding branch are shown by comparing the columns A+B+D and A+D. It is clear that the network performance degrades significantly if either the CE loss has been replaced or the soft-thresholding branch has been removed. Besides, the temporal windowing strategy has shown its advantage in restoring more details when comparing the columns A+B+D and B+D. Specifically, we see the number of abnormal pixels are reduced ((b) and (c)), the needle-like object is revealed ((g) and (h)), and a complete restore of the head of a man has been achieved ((l) and (m)) with the temporal windowing strategy. A clearer outline of the human head, for which even the nose can be distinguished, is restored by the temporal windowing strategy when comparing (q) and (r).


\section{Conclusion}
In this work, we proposed a pixel-wise residual shrinkage network for robust photon-efficient imaging especially in high-noise conditions. Specifically, a noval PRS-block with pixel-wise soft thresholding has been proposed to extract a clean signal from noisy SPAD measurements. By adopting cross-entropy loss for pixel-wise classification, the optimization target becomes more precise, which enforces more confident and centralized depth predictions and thus brings significant performance improvements over the previous works. 

The combination of PRS-block, pixel-wise classification loss and temporal window strategy offers higher imaging quality than state-of-the-art approaches. Extensive experiments on simulated and real-world datasets demonstrate that PRS-Net consistently outperforms the state-of-the-art models by a large margin, for SBRs ranging from 2:10 to 1:100. In particular, PRS-Net shows a superior denoising ability irrespective of SBRs. Besides, benefitting from the simplified architecture of network, PRS-Net requires the shortest processing time and memory in comparison with other deep learning methods. These results suggest an efficient and robust approach for imaging in extreme scenarios such as ultra-long and underwater imaging. Since the experiment of PRS-Net for high-resolution images still takes tens of seconds, an interesting future direction is to further improve the imaging speed for real-time imaging.

\begin{backmatter}

\bmsection{Acknowledgments}
This work was supported by the National Natural Science Foundation of China (Grant No. 62173296) and National Key R\&D Program of China (Grant No. 2018YFB1700100).

\bmsection{Disclosures}
The authors declare that there is no conflict of interest.

\bmsection{Data Availability Statement} Training code, pretrained model and test data are available in Ref. \cite{codeurl}.
\end{backmatter}

\bibliography{references}
\end{document}